# Why is a new Journal of Informetrics needed?[1]

Philipp Mayr*, Walther Umstätter**

* Social Science Information Centre, Lennéstr. 30, 53113 Bonn. Germany
http://www.gesis.org/en/iz/
E-mail: mayr@bonn.iz-soz.de

**Institute for Library and Information Science. Humboldt University. Berlin. Germany
E-mail: h0228kdm@rz.hu-berlin.de

**Abstract**
In our study we analysed 3.889 records which were indexed in the Library and Information Science Abstracts (LISA) database in the research field of informetrics. We can show the core journals of the field via a Bradford (power law) distribution and corroborate on the basis of the restricted LISA data set that it was the appropriate time to found a new specialized journal dedicated to informetrics. According to Bradford's Law of scattering (pure quantitative calculation), Egghe's Journal of Informetrics (JOI) first issue to appear in 2007, comes most probable at the right time.

**Introduction**
In the new Journal of Informetrics (JOI[2]) recent call for papers Leo Egghe mentioned that Elsevier (Oxford, UK) accepted his proposal for the foundation of this new scientific journal. It will be the first journal worldwide which carries 'informetrics' in the journal title. He explained in an editorial why this is necessary (Egghe, 2005). One main reason is the growth of the field informetrics. A broad definition of the field informetrics can be found in Egghe's introductory paper in the first special issue on informetrics in Information Processing and Management Vol. 41(6): "we will use the term "informetrics" as the broad term comprising all-metrics studies related to information science, including bibliometrics (bibliographies, libraries, ...), scientometrics (science policy, citation analysis, research evaluation, ...), webometrics (metrics of the web, the Internet or other social networks such as citation or collaboration networks), ..." (Egghe, 2005 p. 1311).

---

[1] To appear in Cybermetrics. VOLUME 11 (2007): ISSUE 1. PAPER 1
[2] www.elsevier.com/locate/joi

Egghe states several studies to show that the number of papers and authors in informetrics are constantly growing. In terms of multidisciplinarity, the field informetrics is also expanding, this can be shown by new research topics like webometrics and other web based informetric methods (see a review in Björneborn & Ingwersen, 2004).

In the following sections we will show briefly that the distribution of papers published in the broad field of informetrics can be used to consider if a new forum of scientific communication is necessary. For establishing a new scholarly journal publishers generally require, on average, not only hundred papers per year from a hundred scientists (Umstätter, 2003), they also need to have a few hundred interested readers.

**Informetric papers in the LISA database**

To generate a data set of papers in the field of informetrics we used the Library and Information Science Abstracts (LISA) database[3]. The query *bibliomet\* OR cybermet\* OR infomet\* OR informet\* OR scientomet\* OR webomet\** generated a total of 3.889 records that are indexed in LISA (1976-2004, CD-ROM version). The records are indexed under the following terms (Table 1):

| Query terms | Total records | Indexed with LISA descriptors[4] |
|---|---|---|
| bibliomet* | 2851 | 2414 |
| scientomet* | 1631 | 629 |
| informet* | 292 | 90 |
| webomet* | 18 | 11 |
| infomet* | 15 | - |
| cybermet* | 7 | - |

Table 1: Distribution of LISA records in "informetrics"

First we produced a bradfordized list to show the core journals which published the most papers in the field (see Figure 1). By using Bradford's Law of scattering (see e.g. Garfield, 1980; White, 1981; Nicolaisen & Hjørland, 2007 to appear) it is possible to extrapolate the number of publications in the world if we know the core journals. In principle, if a new journal arises the number of scientific publications, has to be about 2 to 3 times higher than 100, because Bradford's Law of scattering means, that all other journals in the world are also attractive for such publications, but with a decrease that follows the power law.

---

[3] http://www.csa.com/factsheets/lisa-set-c.php
[4] The matching LISA descriptors are "Bibliometrics", "Scientometrics", "Informetrics" and "Webometrics".

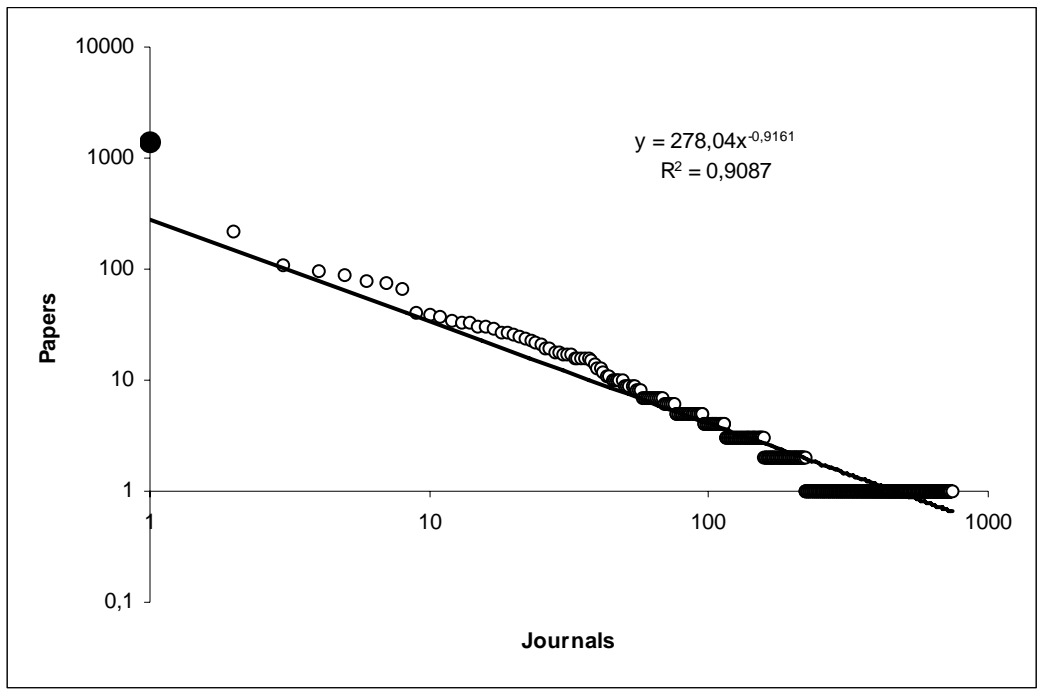

Figure 1: Bradford distribution of papers in "informetrics" (LISA data set)

It is simple to see that the core journal of the field Scientometrics[5] is ~2 to 4 times higher than expected by the Bradford distribution (see Figure 1 and Table 2). The core journals in our Bradford analysis are the following (compare to the journals Egghe mentioned in his editorial (2005)).

| Journal | No. of papers |
|---|---|
| Scientometrics | 1413 |
| Journal of the American Society for Information Science | 218 |
| Nauchno Tekhnicheskaya Informatsiya | 110 |
| Revista Espanola de Documentacion Cientifica | 96 |
| Journal of Information Science | 87 |
| Information Processing and Management | 79 |
| Journal of Documentation | 75 |
| Annals of Library Science and Documentation | 66 |

Table 2: Core journals in "informetrics" (LISA data set)

---

[5] The journal Scientometrics was founded in 1978 and was swiftly included into E. Garfield's Science Citation Index (SCI). It is published at Akadémiai Kiadó, co-published with Springer Science. The Editor-in-Chief was from the outset T. Braun, Hungary. This journal was begun very early when ~100 articles per year appeared in our LISA data set.

It seems to be clear that some journals are highly represented due to a special indexing of LISA. Beyond that we extracted the most productive authors in the informetrics field (see Table 3).

| Author | No. of papers |
|---|---|
| Egghe, L | 64 |
| Glanzel, W | 61 |
| Rousseau, R | 54 |
| Schubert, A | 45 |
| Gupta, B.M | 39 |
| van, Raan, A.F.J | 38 |
| Cronin, B | 26 |
| Garg, K.C | 26 |
| Gomez, I | 24 |
| Small, H | 23 |
| Vinkler, P | 23 |
| Bonitz, M | 21 |
| Braun, T | 21 |

Table 3: Most productive authors (LISA data set)

The distribution of languages shows that the field informetrics is highly dominated by English. The special indexing and coverage of LISA is again a reason for this.

English     81.6%
Russian      4.5%
Spanish      3.3%
Portuguese   1.6%
German       1.5%
Chinese      1.4%
Hungarian    1.0%
Japanese     1.0%

**Growth and competition in informetrics**
Starting with „statistical bibliography", in 1922 we can understand E. Windham Hulm as a pioneer. Interesting works from Bradford (1934), Lotka, Otlet, Pritchard, Zipf and others followed. If we look to the expansion of this field in LISA the growth rate seems to be some 10 years (see Figure 2).

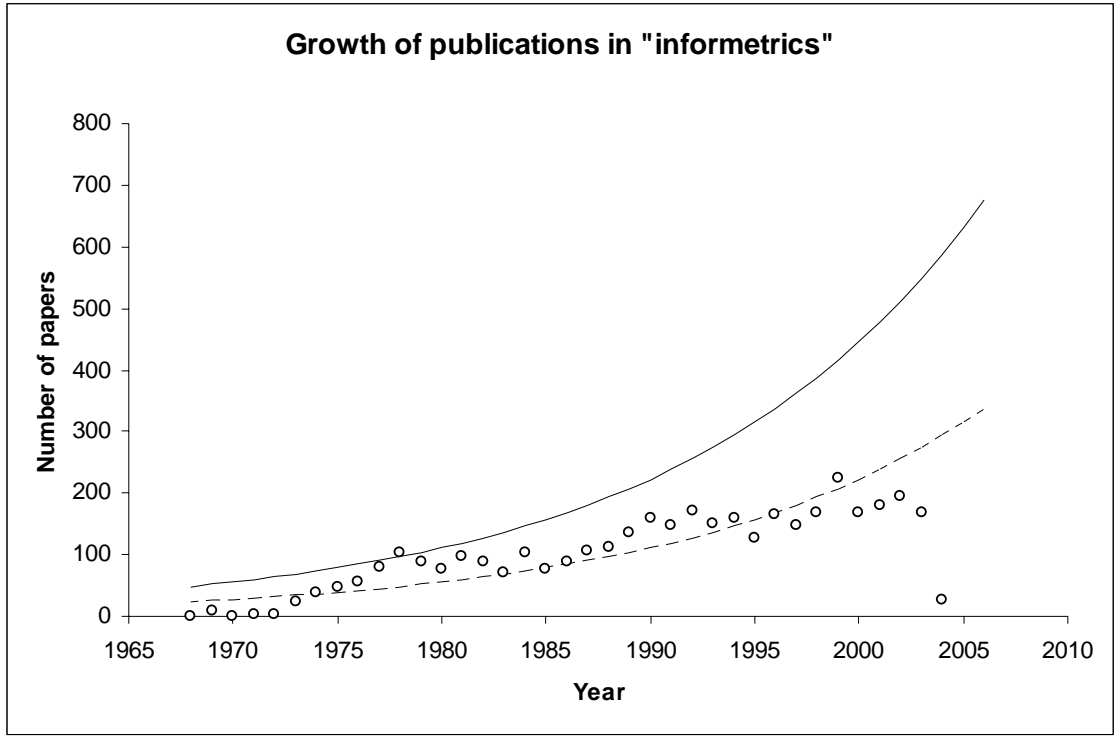

Figure 2: Growth of publications in "informetrics"

A rough estimation of papers in the field of informetrics in LISA shows the doubling rate for a 10 years period to be twice the normally seen value of 20 years. Under the assumption that LISA (dotted line in Figure 2), like similar databases, has only ~40% of the whole amount, we can calculate a number of ~500 published papers in 2002 (straight line in Figure 2).

Calculating the core journals only for the year 2002 (without the comparable too high value of Scientometrics), then we can extrapolate it to the total number of some 100.000 running journal titles in the world (see Figure 3). In this case we get 430 papers in 2002. By a growth rate of 7% per year ($t_2$ = 10 years), we have approximately 600 to 700 papers in 2006. But this value is also based on LISA, and additionally we know that some of the core journals are not indexed adequately.

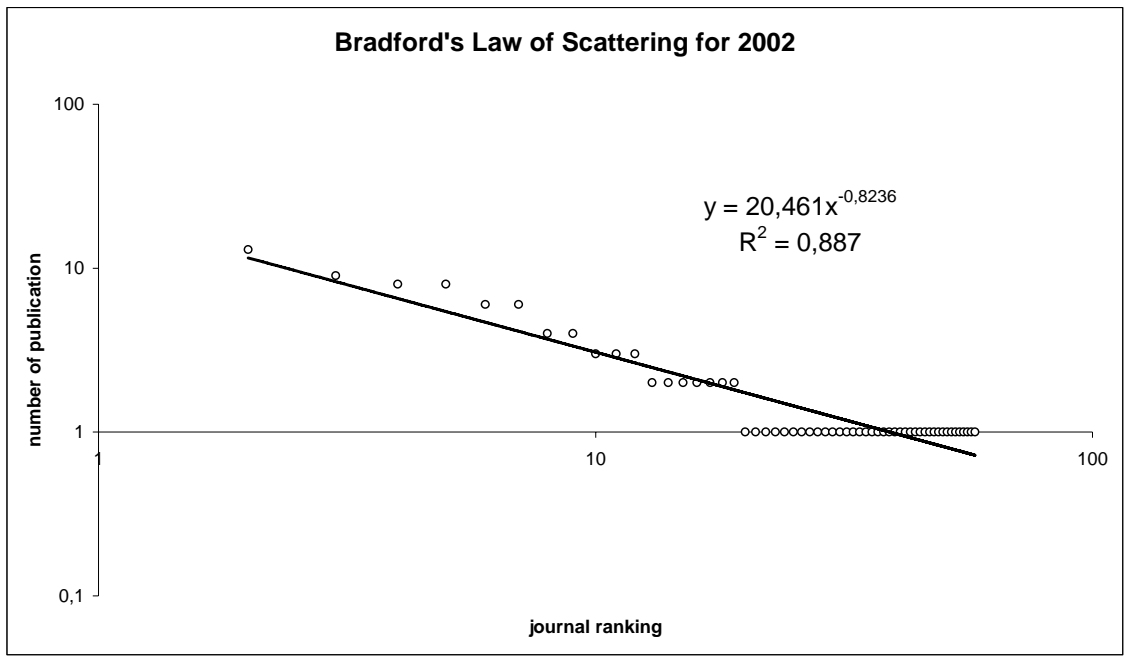

Figure 3: Bradford distribution for 2002 (excluding "Scientometrics")

Another possibility is to calculate Bradford's Law of scattering using as of all the years included in LISA (see Table 4 as an idealized example) to reconstruct the publication growth for all journals with a doubling time of 9.6 years.

| Total no. of papers | No. of source journals |
|---:|---:|
| 750 | 1 |
| 750 | 7 |
| 750 | 49 |
| 750 | 343 |
| 750 | 2.401 |
| 750 | 16.807 |
| 750 | 117.649 |
| Sum of papers 5.250 | |

Table 4: Bradford's Law of scattering using the LISA data set (idealized)

Then we can estimate roughly the number of 1.000 publications in 2005 in this field (see Table 5). The factor 2.5 means, that empirical assessments have shown repeatedly, that the search in one database has only a recall ratio of roughly 40%. Searchers know the motto, since the seventies of the last century, "one database is not enough".

| Year | No. of papers | Papers * factor 2.5 |
|:---:|:---:|---:|
| 2004 | 373 | 932 |
| 2005 | 401 | 1.001 |
| 2006 | 431 | 1.076 |
| 2007 | 463 | 1.157 |

Table 5: An estimation of papers in "informetrics" (idealized)

This can easily be interpreted as meaning that there is room for a new journal parallel to Scientometrics, given the pure quantitative calculation. And it is not too early, especially if it is better or more specialized than the competitors.

It can be noted here that the journal Cybermetrics first appeared in 1997, and surprisingly not one of the papers is indexed in our sample from LISA. Cybermetrics is an "International Journal of Scientometrics, Informetrics and Bibliometrics" dedicated to informetric and webometric studies. It is also an irregular appearing open access journal with only a few papers per year. The editorial board with 29 well known person, is including also Leo Egghe and Ronald Rousseau from Belgium. In 2004 a further journal dedicated to webometrics, named Webology, appeared quarterly as an electronic journal with two issues in 2004, four issues in 2005 and one in 2006. The editor in chief is a PhD candidate from Teheran. It is remarkable that open access journals are not automatically more often indexed and highly cited journals. Webology is cited only 3 times in Web of Science (WoS) with 2 first authors while Cybermetrics was cited 240 times in WoS with 9 first authors with a clear decrease since 1997. More than 50% of the citations go back to papers of R. Rousseau.

A comparison of editorial boards is warranted as theses scientists often use their journals as a platform for their own publications (see Nourmohammadi & Umstätter, 2004) and collaborations[6].

Regardless, new journals are often founded too early. The consequence can be observed very easily by the death of roughly 50% of new journals (Umstätter & Rehm, 1984). The reason for this is also very simple; scientists on the search front always want be the first.

**Conclusion**

There has to be improved standards of quality and increasing specialization has to be found in the growing area of informetrics. Better measurements of information and knowledge would be one of the great desiderata. If we remember that in 1963 a first very rough estimation of the amount of information in the Library of Congress was calculated with $10^{13}$ bit[7].

The new Journal of Informetrics with its focus on "good mathematical (probabilistic) models and explanations of informetric regularities (in the broad sense) and/or papers in which interesting and important data-gathering" comes at the right time, according to our study. The basis of a well known publisher is empirically also very important.

---

[6] Influential Authors in Library and Information Science 2000-2002, see http://www.umu.se/inforsk/LIS/LIS2000-2002.htm by Olle Persson.

[7] Weinberg-Kommission: Science, Government and Information. Report of The President's Science Advisory Commitee USA. Washington (1963). It should be possible also to calculate the knowledge (Umstätter, 1998) of libraries in bit, as a much more compressed form of evidence based information.